\documentclass[aps,prb,twocolumn,10pt,showpacs,amsmath,amssymb]{revtex4-1}
\usepackage{graphicx}
\begin{document}
\title{$^{119}$Sn-NMR investigations on superconducting Ca$_3$Ir$_4$Sn$_{13}$: Evidence for multigap superconductivity}
\author{R. Sarkar}
\altaffiliation[]{rajibsarkarsinp@gmail.com}%Lines break automatically or can be forced with \\
\author{F. Br\"uckner}
\author{M. G\"unther}
\affiliation{Institute for Solid State Physics, TU Dresden, D-01069 Dresden, Germany}
\author{C. Petrovic}
\author{Kefeng Wang}
\affiliation{Condensed Matter Physics and Materials Science Department, Brookhaven National Laboratory, Upton, NY-11973, USA}
\author{P. K. Biswas}
\author{H. Luetkens}
\author{E. Morenzoni}
\author{A. Amato}
\affiliation{Laboratory for Muon Spin Spectroscopy, Paul Scherrer Institute, CH-5232 Villigen PSI, Switzerland}
\author{H-H. Klauss}
\affiliation{Institute for Solid State Physics, TU Dresden, D-01069 Dresden, Germany}

%\author[]{R. Sarkar \footnote{Author to whom all correspondence should be made.}, F. Br\"uckner, M. G\"unter, \dag\ C. Petrovic, K. Wang, \S\ H. Luetkens, P. K. Biswas, \S\ E. Morenzoni, \S\ A. Amato, H-H. Klauss}
%\address{{Institute of Solid State Physics, TU Dresden, D-01069 Dresden, Germany}}
%\address{\dag {Condensed Matter Physics and Materials Science Department, Brookhaven National Laboratory, Upton, NY-11973, USA}}
%\address{\S {Laboratory for Muon Spin Spectroscopy, Paul Scherrer Institute, CH-5232 Villigen PSI, Switzerland}}
%\ead{rajibsarkarsinp@gmail.com}

%\affiliation{% Max-Planck Institute for Chemical Physics of Solids, 01187 Dresden, Germany\\
%Max-Planck Institute for Chemical Physics of Solids, 01187
%Dresden, Germany}
%\author{H. Luetkens}
%\affiliation{Laboratory for Muon-Spin Spectroscopy, Paul-Scherrer-Institute, CH-5232 Villigen, Switzerland}

\date{\today }

\begin{abstract}
We report bulk superconductivity (SC) in Ca$_3$Ir$_4$Sn$_{13}$ by means of $^{119}$Sn nuclear magnetic resonance (NMR) experiments. Two classical signatures of BCS superconductivity in spin-lattice relaxation rate ($1/T_1$), namely the Hebel-Slichter coherence peak just below the $T_c$ and the exponential decay in the superconducting phase, are evident. The noticeable decrease of $^{119}$Sn Knight shift below $T_c$ indicates spin-singlet superconductivity. The temperature dependence of the spin-lattice relaxation rate ($1/T_1$) is convincingly described by the multigap isotropic superconducting gap. Present NMR experiments do not witness any sign of enhanced spin fluctuations in the normal state. 
\end{abstract}

%\keywords{Superconductivity}
%\pacs{74.70.-b, 76.60.-k, 75.30.Fv, 74.25.Jb}

\maketitle

\section{\textbf{Introduction}}
The interplay between superconductivity (SC), charge and spin fluctuations is the central topic of 
research in the field of unconventional SC. \cite{{Kim-2008},{sun-2008},{Dagotto2005},{Morosan2006},{Mathur1998},{Monthoux2007}} Ca$_3$Ir$_4$Sn$_{13}$ is a member of the material class of 
superconducting and/or magnetic ternary intermetallic compounds.\cite{Remeika1980923} While Ca$_3$Ir$_4$Sn$_{13}$ was first synthesized more than
30 years ago, very recently it received revived attention in the condensed matter community because of its interesting 
physical properties. \cite{Espinosa1980} 
%In early studies, it was reported that the superconducting transition at $T_c$\,=\,7\,K along with the Quasiskutterudite crystal structure. 
In early studies superconducting transition at $T_c$\,=\,7\,K and quasiskutteridite crystal structure were reported.
 More recent thermodynamic and transport measurements indicate an anomaly 
at $T^*$~=~38\,K, well above the superconducting transition. \cite{yang2010} Yang  first proposed that the anomaly $et$ $al.$
is the result of ferromagnetic spin fluctuations coupled to SC. \cite{yang2010} In contrast $\mu$SR experiments do not find any 
evidence of enhanced spin fluctuations in the $\mu$SR time scale. \cite{Gerber-2013} On the other hand Wang $et$ $al.$ attributed that
anomaly to a significant Fermi surface reconstruction and the opening of a charge density wave gap at the super-lattice transition. \cite{{Petrovic2012},{Klintberg2012}} While they classified Ca$_3$Ir$_4$Sn$_{13}$ as a weakly correlated nodeless superconductor, recent $\mu$SR investigations reveal that the electron-phonon pairing interaction is in the strong-coupling limit. Furthermore $\mu$SR and macroscopic measurements indicated the presence of multiple isotropic gaps with different magnitudes. \cite{{Gerber-2013},{Zhou-2012}} However so far from microscopic viewpoint superconducting gap structure and the possible presence of the spin fluctuations are not completely understood in Ca$_3$Ir$_4$Sn$_{13}$.   
%Ca$_3$Ir$_4$Sn$_{13}$ was found to exhibit superconducting transition with $T_c \approx$~7 K. It received considerable attention due to the
%possible coexistence of superconductivity and ferromagnetic spin fluctuation as well as the three-dimensional charge density wave (CDW) from the super-lattice transition.  While thermal, transport, and thermodynamic characterization of Ca$_3$Ir$_4$Sn$_{13}$ single crystals suggest that it is a weakly correlated nodeless
%superconductor, recent $\mu$SR investigation reveals that the electron-phonon pairing interaction is in the strong-coupling limit. 
\\
\indent
In order to investigate the wave symmetry of the superconducting order parameter and to prove/disprove the presence of ferromagnetic spin fluctuations and its possible impact on superconductivity, we have carried out detailed NMR experiments. Here we present $^{119}$Sn NMR investigations on Ca$_3$Ir$_4$Sn$_{13}$ polycrystalline samples. We discuss the symmetry of the superconducting order parameter together with the normal state properties. The present results of spin-lattice relaxation rate ($1/T_1$) indicate that Ca$_3$Ir$_4$Sn$_{13}$ is a BCS superconductor. While the Arrhenius plot of $1/T_1$ vs $1/T$ gives a superconducting gap value of $|2\Delta/k_BT_c|$~=~4.4, the temperature dependence of the $1/T_1$ can be described by the multiple isotropic gaps. This multigap fit gives the gap values of $\frac{2\Delta_1(0)}{k_B T_c}=7$ and $\frac{2\Delta_1(0)}{k_B T_c}=1.5$, respectively. So far NMR experiments do not find any evidence of enhanced spin fluctuations. Our investigations are in good agreement with the reported $\mu$SR studies.\cite{Gerber-2013} 

\section{\textbf{Experimental details}}
Single crystals of Ca$_3$Ir$_4$Sn$_{13}$ were grown and characterized as described elsewhere. \cite{Petrovic2012} The NMR measurements were carried out using the conventional pulsed NMR technique on $^{119}$Sn (nuclear
spin $I=1/2$)
% and gyromagnetic ratio $\gamma/2\pi~=~7.2914$ MHz/T)
nuclei in a temperature range 1.5\,K $\leq$ $T$ $\leq $ $60$\,K at 33 MHz in random oriented polycrystalline sample. For this
purpose we have crushed the single crystal into powder and mixed them in paraffin. The field sweep
NMR spectra were obtained by integrating the spin-echo in the time domain and plotting the resulting intensity as a function of
the field. The spin lattice relaxation rate ($1/T_{1}$) experiments were performed at 33 MHz following the standard saturation
recovery method.

\section{\textbf{Results and Discussion}}
\subsection{\label{sec:level2}Sn NMR Spectra and Shift}
\begin{figure}
\centering
\includegraphics[scale=0.94]{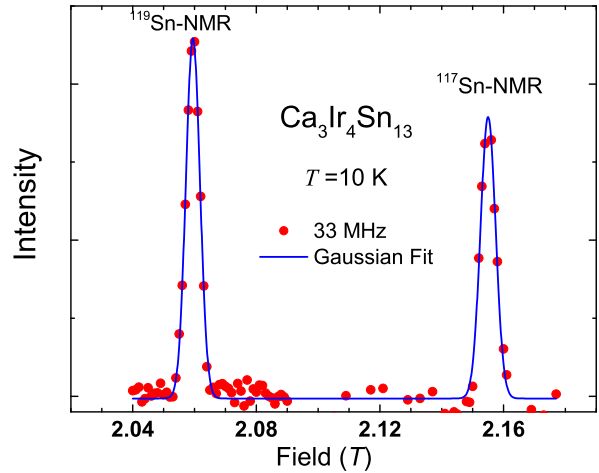}
\caption{\label{fig1} $^{119/117}$Sn NMR field sweep spectrum taken at $T$~=~10 K and 33 MHz. Line indicates multi peak Gaussian fit.}
\end{figure}

\begin{figure}
\centering
\includegraphics[scale=.94]{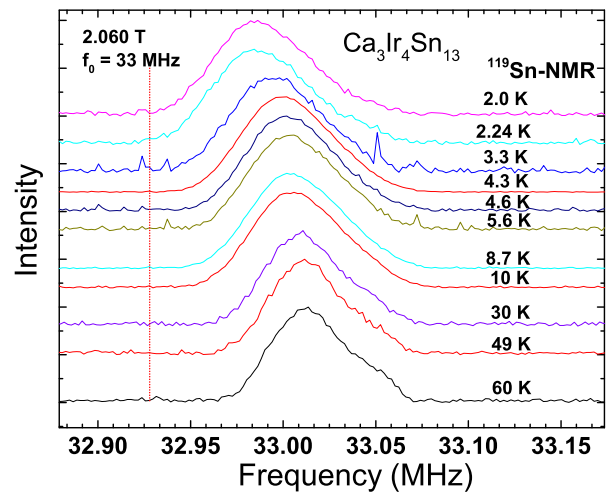}
\caption{\label{fig2} Temperature dependence of the $^{119}$Sn spectra. vertical dotted line indicate the dia-magnetic reference position.}
\end{figure}

Sn-field sweep NMR spectra are shown in Fig.~\ref{fig1} taken at a frequency of 33 MHz and at a temperature of 10 K. Two Sn isotopes namely $^{119}$Sn and $^{117}$Sn in agreement with their respective textbook gamma values and natural abundances are clearly resolved. In the following we will focus on the $^{119}$Sn NMR investigations. 
Figure~\ref{fig2} shows the temperature dependence of the $^{119}$Sn Fourier transformed NMR spectra. It represents a single isotropic line as expected for $I$~=~1/2 nuclei in cubic crystal structure. Even if for random polycrystalline sample the full width of half maximum (FWHM) is found to be 55 kHz indicating very good sample quality. With varying temperature spectral shape doesn't change significantly, except those are shifted with lowering the temperature towards the low frequency side prominently in the superconducting state. Just below the $T_c$ spectra are shifted considerably.
% indicating change of Shift. 
 The Knight shift was determined straightforwardly from the peak of the spectra.
The estimated Knight shift ($^{119}K(\%))$ as a function of temperature is plotted in Fig.~\ref{fig3} revealing a significant drop at $T_c$. The NMR shift gives the information of the local susceptibility. For a spin singlet superconductor it decreases in the superconducting state. In the normal state the Knight shift is constant without any anomaly around 38 K (Fig.~\ref{fig3}) suggesting simple metallic state of Ca$_3$Ir$_4$Sn$_{13}$. A small change is observed  around 20 K (Fig.~\ref{fig3}). To understand the origin of this small change around 20 K, we have also performed at low field (1 T) the $^{119}$Sn Knight shift measurements however no changes could be detected in this temperature. This rules out the possibilities of intrinsic nature of this small change. $^{119}$Sn Knight shift decreases in the superconducting state below 4.5\,K. Even though no orbital contribution has been subtracted, the significant decrease of Knight shift in the superconducting state indicates that Ca$_3$Ir$_4$Sn$_{13}$ is a spin-singlet superconductor.    

\subsection{\label{sec:level2}$^{119}$Sn NMR spin-lattice relaxation rate $^{119}(1/T_1$)}
\begin{figure}
\centering
\includegraphics[scale=.94]{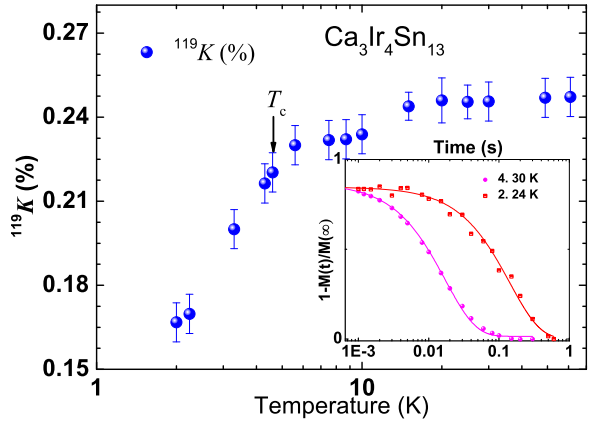}
\caption{\label{fig3} (main panel) Temperature dependence of the Knight shift $^{119}$K measured at 33 MHz. Inset shows the recovery curves for the spin lattice relaxation rate ($1/T_1$) measurements.}
\end{figure}

\begin{figure}
\centering
\includegraphics[scale=0.9]{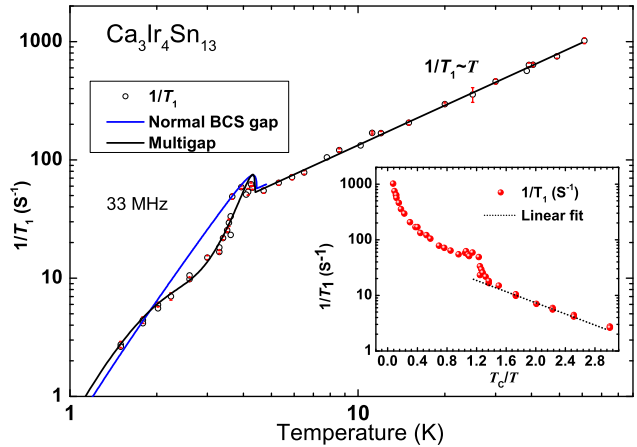}
\caption{\label{fig4} (main panel )$^{119}$Sn spin-lattice relaxation rate ($1/T_1$) vs $T$ plot at frequency 33 MHz and magnetic field 2.06 T  for Ca$_3$Ir$_4$Sn$_{13}$. Lines represent different theoretical model as described in the text. Inset shows the $(1/T_1$) vs $1/T$ plot, commonly known as Arrhenius plot, together with the linear fit. }
\end{figure}

To investigate the wave symmetry of the superconducting order parameter, we have carried out NMR spin-lattice relaxation rate $^{119}(1/T_1$) experiments. 
%Spin-lattice relaxation rate $1/T_{1}$ was measured by the standard saturation recovery method. 
Spin-lattice relaxation rate $^{119}(1/T_1$) recovery curves could be fitted consistently with a single $T_{1}$ component using single exponential in the whole temperature range suggesting a very homogeneous system,
\begin{equation}
1-\frac{M(t)}{M(\infty)}~=~Ae^{-t/T_{1}},
\label{exp}
\end{equation}
where $M(t)$ is the nuclear magnetization at a time $t$ after the
saturation pulse and $M(\infty)$ is the equilibrium magnetization. In the inset of the Fig.~\ref{fig3} two recovery curves in the superconducting state are shown.
\\
\indent
The temperature dependence of $^{119}(1/T_1$) is presented in Fig.~\ref{fig4}. In the temperature range 4.5~-~60\,K, $1/T_{1}$ follows a linear relation with temperature indicating the validity of the Korringa law and implying a simple metallic state for Ca$_3$Ir$_4$Sn$_{13}$. However with lowering the temperature, just below $T$~=~4.5\,K, $^{119}(1/T_1$) displays a Hebel-Slichter coherence peak and followed by this $^{119}(1/T_1$) sharply decreases with two different slopes. To understand the superconducting ordered state and to get a first educated guess of superconducting order parameters, as a first approach we made an Arrhenius plot, which is depicted in the inset of Fig.~\ref{fig4}.\cite{Harada-JPSJ-2007} This plot clearly shows two distinct slopes in the superconducting regime. Well below $T_c$ in the temperature range 3.3~-~1.5 \,K a clear exponential decay is realized. The dotted line is the description of the linear fit with the following equation: 
\begin{equation}
\ln(1/T_1)=\ln A-\Delta/k_BT.
\label{linearfit}
\end{equation}
The estimated fit parameter values are $\ln$~A~=~1.89 and $2|\Delta/k_BT|$~$\approx$~4.4, respectively. The estimated gap value is bit higher than the value of the conventional BCS gap value, 2$\Delta/k_B T_c$~=~3.5. So far two classical signatures of BCS superconductivity: 1) Hebel-Slichter coherence peak in $1/T_1$ vs $T$ plot just below $T_c$ and 2) exponential decay in $1/T_1$ vs $T$ plot are evident reflecting the BCS character of superconductivity in Ca$_3$Ir$_4$Sn$_{13}$.
%Now we put an effort to model the experimental $1/T_1$ data  by using a full BCS theory. To do that we have used the generalized Hebel-Slichter formula \cite{MacLaughlin-1976} 
%\begin{equation}
%T_{1N}/T_{1S}=2/k_BT\int_{0}^{\propto}(N_s(E)N_s(E^\prime)+ M_s(E)M_s(E^\prime))\times f(E)(1-f(E^\prime))dE
%\label{Hebel-Slichter1}
%\end{equation}
%with the density of states $N_s$ and anomalous density of states $M_s$ 

%\begin{equation}
%N_s(E)= \Re\R\left[ \int P(a) \frac {E}{\sqrt{E^2-\Delta^{2}_{0}|1+a|^2}}da\right],
%\label{Hebel-Slichter2}
%\end{equation}
%\begin{equation}
%M_s(E)= \Re\R\left[ \int P(a) \frac {\Delta_0(1+a)}{\sqrt{E^2-\Delta^{2}_{0}|1+a|^2}}da\right],
%\label{Hebel-Slichter2}
%\end{equation}
%where $f$ is the Fermi function and $E^{\prime}$ = E+$\hbar\omega_{nuc}$. $P(a)$ is the distribution of the anisotropic gap 
%$\Delta(\Omega)~=~\Delta_0 (1+a(\Omega))$. $P(a) = \delta(a)$ describes the isotropic BCS case. The superconducting gap is estimated by solving the full BCS gap equation numerically for a specified $T_c$ and 
In the following we shall discuss more deeper analysis of the spin-lattice relaxation rate ($1/T_1$) data. In order to achieve that we perform two fits of the spin-lattice relaxation rate. The first represents a simple $s$-wave symmetry with a phenomenological distribution of gaps as described in Ref.\cite{maclaughlin76} and can be interpreted as an anisotropy of the order parameter. The distribution is set to be a rectangular function in the interval $[\Delta_0(1-\delta/2),\Delta_0(1+\delta/2)]$. The anisotropy broadens the peak in the density of states. Because this model deviates significantly from the data in Fig.~\ref{fig4}, we include a multigap $s$-wave model with two gaps $\Delta_1$ and $\Delta_2$.\cite{{Gerber-2013},{Zhou-2012}} 
%We have followed this approach as we motivated by the prediction of the multigap superconductivity as reported in the literature. 
 The temperature dependence of the gap is calculated by solving the coupled gap equations as suggested by Suhl $et$ $al.$ \cite{suhl59} Then the relaxation rate $1/T_1$ can be evaluated with the integral
\begin{eqnarray}
\frac{T_{1\mathrm{N}}}{T_{1\mathrm{S}}}=\frac{2}{k_\mathrm{B} T} \int\limits_{0}^{\infty} \sum_{n=1,2} c_{n} \left\lbrace N_\mathrm{s}^{n}(E)^{2} + M_\mathrm{s}^{n}(E)^{2} \right\rbrace \nonumber \\  f(E) \left(1-f(E)\right) dE, \label{HSequ}
\end{eqnarray}
with the anomalous density of states  $N_{s}^{n}(E)$ and $M_{s}^{n}(E)$ of the band with index $n$, the relative weight of the density of states $c_n$ and the Fermi function $f$. This equation implies that inter-band scattering of the electrons within the relaxation process is neglected, which is motivated by the weak inter-band scattering channels necessary for significant multigap effects. The anisotropy of the particular band is covered in the same manner as in the single gap model.

In the Fig.~\ref{fig4} blue and black lines represent the normal BCS and multigap fits, respectively. Important parameters as obtained from the multigap fit are as follows
%\begin{align}
%$\delta_1$~=~0.6, $\delta_1$~=~0.6, 
$c_1$~=~0.83, $c_2$~=~0.17, $\frac{2\Delta_1(0)}{k_B T_c}$~=~7 and $\frac{2\Delta_1(0)}{k_B T_c}=1.5$. One of the estimated gap value is $\frac{2\Delta_1(0)}{k_B T_c}$~=~7 which is much larger than the weak-coupling  BCS value of $\frac{2\Delta}{k_B T_c}$~=~3.5, suggesting a strong electron-phonon coupling and being consistent with the $\mu$SR results. \cite{Gerber-2013} 
\\
\indent
% Now we discuss the presence of ferromagnetic fluctuations and it's possible coupling to the superconductivity in Ca$_3$Ir$_4$Sn$_{13}$. However,
In the present local probe NMR investigations neither the Knight shift (static susceptibility) nor spin-lattice relaxation rate ($1/T_1$) (dynamic susceptibility) suggests the presence of enhanced spin fluctuations. In contrast NMR experiments demonstrate the simple metallic nature of Ca$_3$Ir$_4$Sn$_{13}$.     
%\end{align}

%\textit{Next we move our discussion on to prove or disprove the presence of Ferromagtic correlation and/or fluctuations in Ca$_3$Ir$_4$Sn$_{13}$ as proposed by Yang et al. \cite{yang2010} For free-electron metals the Korringa relation is given by $1/T_1TK^2=S_0=\pi\hbar\gamma_N^2k_B/\mu_B^2$, where $\gamma_N$ is the nuclear gyromagnetic ratio of $^{119}$Sn and $k_B$ is the Boltzmann constant. Including electronic correlations leads to the modified Korringa relation  $1/T_1TK^2=S=\alpha S_0$. The so-called Korringa product $(1/T_1TK^2S_0)=\alpha $ is a very useful probe for correlations, where $\alpha=1$ indicates the absence of correlations, $\alpha > 1$ indicates an antiferromagnetic correlation and $\alpha < 1$ indicates ferromagnetic correlation. While for the $^{119}$Sn nuclei $S_0$ is 5.4 $\times$ 10$^5$ 1/sK, experimentally we observed 2.37$\times$ 10$^6$ 1/sK at 60 K. This gives $\alpha$ value 4.4 indicating the absence of ferromagnetic correlation.} 

\section{Conclusions}
To conclude $^{119}$Sn NMR experiments on Ca$_3$Ir$_4$Sn$_{13}$ were carried out. The $^{119}$Sn Knight shift has revealed a significant decrease below $T_c$, suggesting a spin-singlet superconductivity. Two classical signatures of BCS superconductors, namely 1) the Hebel-Slichter coherence peak in $1/T_1$ vs. $T$ just below $T_c$, and 2) exponential decay in $1/T_1$ are evident. By simple Arrhenius plot, the estimated gap value is  $|2\Delta/k_BT_c|$~=~ 4.4, which is higher than the BCS value. The temperature dependence of $1/T_1$ data in the superconducting state could be described by the multigap superconductivity in Ca$_3$Ir$_4$Sn$_{13}$ which exhibiting gap values of $\frac{2\Delta_1(0)}{k_B T_c}=7$ and $\frac{2\Delta_1(0)}{k_B T_c}=1.5$, respectively. From NMR point of view Ca$_3$Ir$_4$Sn$_{13}$ does not have any significant enhanced spin fluctuations in the normal state.
% So we can safely disregard the idea of spin fluctuation in Ca$_3$Ir$_4$Sn$_{13}$.      

\section*{Acknowledgments} R. Sarkar is thankful to DFG for the financial support through grant No. SA 2426/1-1. Work at Brookhaven is supported by the US DOE under
Contract No. DE-AC02-98CH10886.
%\end{acknowledgments}

%\section*{References}
\bibliographystyle{apsrev}
\bibliography{Sarkar}

\begin{thebibliography}{16}
\expandafter\ifx\csname natexlab\endcsname\relax\def\natexlab#1{#1}\fi
\expandafter\ifx\csname bibnamefont\endcsname\relax
  \def\bibnamefont#1{#1}\fi
\expandafter\ifx\csname bibfnamefont\endcsname\relax
  \def\bibfnamefont#1{#1}\fi
\expandafter\ifx\csname citenamefont\endcsname\relax
  \def\citenamefont#1{#1}\fi
\expandafter\ifx\csname url\endcsname\relax
  \def\url#1{\texttt{#1}}\fi
\expandafter\ifx\csname urlprefix\endcsname\relax\def\urlprefix{URL }\fi
\providecommand{\bibinfo}[2]{#2}
\providecommand{\eprint}[2][]{\url{#2}}

\bibitem[{\citenamefont{Kim et~al.}(2010)\citenamefont{Kim, Barath, Chen, Joe,
  Fradkin, Abbamonte, and Cooper}}]{Kim-2008}
\bibinfo{author}{\bibfnamefont{M.}~\bibnamefont{Kim}},
  \bibinfo{author}{\bibfnamefont{H.}~\bibnamefont{Barath}},
  \bibinfo{author}{\bibfnamefont{X.}~\bibnamefont{Chen}},
  \bibinfo{author}{\bibfnamefont{Y.-I.} \bibnamefont{Joe}},
  \bibinfo{author}{\bibfnamefont{E.}~\bibnamefont{Fradkin}},
  \bibinfo{author}{\bibfnamefont{P.}~\bibnamefont{Abbamonte}},
  \bibnamefont{and} \bibinfo{author}{\bibfnamefont{S.~L.}
  \bibnamefont{Cooper}}, \bibinfo{journal}{Advanced Materials}
  \textbf{\bibinfo{volume}{22}}, \bibinfo{pages}{1148} (\bibinfo{year}{2010}),
  ISSN \bibinfo{issn}{1521-4095},
  \urlprefix\url{http://dx.doi.org/10.1002/adma.200904246}.

\bibitem[{\citenamefont{Sun et~al.}(2008)\citenamefont{Sun, Fregoso, Lawler,
  and Fradkin}}]{sun-2008}
\bibinfo{author}{\bibfnamefont{K.}~\bibnamefont{Sun}},
  \bibinfo{author}{\bibfnamefont{B.~M.} \bibnamefont{Fregoso}},
  \bibinfo{author}{\bibfnamefont{M.~J.} \bibnamefont{Lawler}},
  \bibnamefont{and} \bibinfo{author}{\bibfnamefont{E.}~\bibnamefont{Fradkin}},
  \bibinfo{journal}{Phys. Rev. B} \textbf{\bibinfo{volume}{78}},
  \bibinfo{pages}{085124} (\bibinfo{year}{2008}),
  \urlprefix\url{http://link.aps.org/doi/10.1103/PhysRevB.78.085124}.

\bibitem[{\citenamefont{Dagotto}(2005)}]{Dagotto2005}
\bibinfo{author}{\bibfnamefont{E.}~\bibnamefont{Dagotto}},
  \bibinfo{journal}{Science} \textbf{\bibinfo{volume}{309}},
  \bibinfo{pages}{257} (\bibinfo{year}{2005}).

\bibitem[{\citenamefont{Morosan et~al.}(2006)\citenamefont{Morosan, Zandbergen,
  Dennis, Bos, Onose, Klimczuk, Ramirez, Ong, and Cava}}]{Morosan2006}
\bibinfo{author}{\bibfnamefont{E.}~\bibnamefont{Morosan}},
  \bibinfo{author}{\bibfnamefont{H.~W.} \bibnamefont{Zandbergen}},
  \bibinfo{author}{\bibfnamefont{B.~S.} \bibnamefont{Dennis}},
  \bibinfo{author}{\bibfnamefont{J.~W.~G.} \bibnamefont{Bos}},
  \bibinfo{author}{\bibfnamefont{Y.}~\bibnamefont{Onose}},
  \bibinfo{author}{\bibfnamefont{T.}~\bibnamefont{Klimczuk}},
  \bibinfo{author}{\bibfnamefont{A.~P.} \bibnamefont{Ramirez}},
  \bibinfo{author}{\bibfnamefont{N.~P.} \bibnamefont{Ong}}, \bibnamefont{and}
  \bibinfo{author}{\bibfnamefont{R.~J.} \bibnamefont{Cava}},
  \bibinfo{journal}{Nat Phys} \textbf{\bibinfo{volume}{2}},
  \bibinfo{pages}{544} (\bibinfo{year}{2006}),
  \urlprefix\url{http://dx.doi.org/10.1038/nphys360}.

\bibitem[{\citenamefont{Mathur et~al.}(1998)\citenamefont{Mathur, Grosche,
  Julian, Walker, Freye, Haselwimmer, and Lonzarich}}]{Mathur1998}
\bibinfo{author}{\bibfnamefont{N.~D.} \bibnamefont{Mathur}},
  \bibinfo{author}{\bibfnamefont{F.~M.} \bibnamefont{Grosche}},
  \bibinfo{author}{\bibfnamefont{S.~R.} \bibnamefont{Julian}},
  \bibinfo{author}{\bibfnamefont{I.~R.} \bibnamefont{Walker}},
  \bibinfo{author}{\bibfnamefont{D.~M.} \bibnamefont{Freye}},
  \bibinfo{author}{\bibfnamefont{R.~K.~W.} \bibnamefont{Haselwimmer}},
  \bibnamefont{and} \bibinfo{author}{\bibfnamefont{G.~G.}
  \bibnamefont{Lonzarich}}, \bibinfo{journal}{Nature}
  \textbf{\bibinfo{volume}{394}}, \bibinfo{pages}{39} (\bibinfo{year}{1998}),
  \urlprefix\url{http://dx.doi.org/10.1038/27838}.

\bibitem[{\citenamefont{Monthoux et~al.}(2007)\citenamefont{Monthoux, Pines,
  and Lonzarich}}]{Monthoux2007}
\bibinfo{author}{\bibfnamefont{P.}~\bibnamefont{Monthoux}},
  \bibinfo{author}{\bibfnamefont{D.}~\bibnamefont{Pines}}, \bibnamefont{and}
  \bibinfo{author}{\bibfnamefont{G.~G.} \bibnamefont{Lonzarich}},
  \bibinfo{journal}{Nature} \textbf{\bibinfo{volume}{450}},
  \bibinfo{pages}{1177} (\bibinfo{year}{2007}),
  \urlprefix\url{http://dx.doi.org/10.1038/nature06480}.

\bibitem[{\citenamefont{Remeika et~al.}(1980)\citenamefont{Remeika, Espinosa,
  Cooper, Barz, Rowell, McWhan, Vandenberg, Moncton, Fisk, Woolf
  et~al.}}]{Remeika1980923}
\bibinfo{author}{\bibfnamefont{J.}~\bibnamefont{Remeika}},
  \bibinfo{author}{\bibfnamefont{G.}~\bibnamefont{Espinosa}},
  \bibinfo{author}{\bibfnamefont{A.}~\bibnamefont{Cooper}},
  \bibinfo{author}{\bibfnamefont{H.}~\bibnamefont{Barz}},
  \bibinfo{author}{\bibfnamefont{J.}~\bibnamefont{Rowell}},
  \bibinfo{author}{\bibfnamefont{D.}~\bibnamefont{McWhan}},
  \bibinfo{author}{\bibfnamefont{J.}~\bibnamefont{Vandenberg}},
  \bibinfo{author}{\bibfnamefont{D.}~\bibnamefont{Moncton}},
  \bibinfo{author}{\bibfnamefont{Z.}~\bibnamefont{Fisk}},
  \bibinfo{author}{\bibfnamefont{L.}~\bibnamefont{Woolf}},
  \bibnamefont{et~al.}, \bibinfo{journal}{Solid State Communications}
  \textbf{\bibinfo{volume}{34}}, \bibinfo{pages}{923 } (\bibinfo{year}{1980}),
  ISSN \bibinfo{issn}{0038-1098},
  \urlprefix\url{http://www.sciencedirect.com/science/article/pii/0038109880910996}.

\bibitem[{\citenamefont{Espinosa}(1980)}]{Espinosa1980}
\bibinfo{author}{\bibfnamefont{G.}~\bibnamefont{Espinosa}},
  \bibinfo{journal}{Materials Research Bulletin} \textbf{\bibinfo{volume}{15}},
  \bibinfo{pages}{791 } (\bibinfo{year}{1980}), ISSN \bibinfo{issn}{0025-5408},
  \urlprefix\url{http://www.sciencedirect.com/science/article/pii/0025540880900136}.

\bibitem[{\citenamefont{Yang et~al.}(2010)\citenamefont{Yang, Chen, Michioka,
  and Yoshimura}}]{yang2010}
\bibinfo{author}{\bibfnamefont{J.}~\bibnamefont{Yang}},
  \bibinfo{author}{\bibfnamefont{B.}~\bibnamefont{Chen}},
  \bibinfo{author}{\bibfnamefont{C.}~\bibnamefont{Michioka}}, \bibnamefont{and}
  \bibinfo{author}{\bibfnamefont{K.}~\bibnamefont{Yoshimura}},
  \bibinfo{journal}{Journal of the Physical Society of Japan}
  \textbf{\bibinfo{volume}{79}}, \bibinfo{pages}{113705}
  (\bibinfo{year}{2010}), \eprint{http://dx.doi.org/10.1143/JPSJ.79.113705},
  \urlprefix\url{http://dx.doi.org/10.1143/JPSJ.79.113705}.

\bibitem[{\citenamefont{Gerber et~al.}(2013)\citenamefont{Gerber, Gavilano,
  Medarde, Pomjakushin, Baines, Pomjakushina, Conder, and
  Kenzelmann}}]{Gerber-2013}
\bibinfo{author}{\bibfnamefont{S.}~\bibnamefont{Gerber}},
  \bibinfo{author}{\bibfnamefont{J.~L.} \bibnamefont{Gavilano}},
  \bibinfo{author}{\bibfnamefont{M.}~\bibnamefont{Medarde}},
  \bibinfo{author}{\bibfnamefont{V.}~\bibnamefont{Pomjakushin}},
  \bibinfo{author}{\bibfnamefont{C.}~\bibnamefont{Baines}},
  \bibinfo{author}{\bibfnamefont{E.}~\bibnamefont{Pomjakushina}},
  \bibinfo{author}{\bibfnamefont{K.}~\bibnamefont{Conder}}, \bibnamefont{and}
  \bibinfo{author}{\bibfnamefont{M.}~\bibnamefont{Kenzelmann}},
  \bibinfo{journal}{Phys. Rev. B} \textbf{\bibinfo{volume}{88}},
  \bibinfo{pages}{104505} (\bibinfo{year}{2013}),
  \urlprefix\url{http://link.aps.org/doi/10.1103/PhysRevB.88.104505}.

\bibitem[{\citenamefont{Wang and Petrovic}(2012)}]{Petrovic2012}
\bibinfo{author}{\bibfnamefont{K.}~\bibnamefont{Wang}} \bibnamefont{and}
  \bibinfo{author}{\bibfnamefont{C.}~\bibnamefont{Petrovic}},
  \bibinfo{journal}{Phys. Rev. B} \textbf{\bibinfo{volume}{86}},
  \bibinfo{pages}{024522} (\bibinfo{year}{2012}),
  \urlprefix\url{http://link.aps.org/doi/10.1103/PhysRevB.86.024522}.

\bibitem[{\citenamefont{Klintberg et~al.}(2012)\citenamefont{Klintberg, Goh,
  Alireza, Saines, Tompsett, Logg, Yang, Chen, Yoshimura, and
  Grosche}}]{Klintberg2012}
\bibinfo{author}{\bibfnamefont{L.~E.} \bibnamefont{Klintberg}},
  \bibinfo{author}{\bibfnamefont{S.~K.} \bibnamefont{Goh}},
  \bibinfo{author}{\bibfnamefont{P.~L.} \bibnamefont{Alireza}},
  \bibinfo{author}{\bibfnamefont{P.~J.} \bibnamefont{Saines}},
  \bibinfo{author}{\bibfnamefont{D.~A.} \bibnamefont{Tompsett}},
  \bibinfo{author}{\bibfnamefont{P.~W.} \bibnamefont{Logg}},
  \bibinfo{author}{\bibfnamefont{J.}~\bibnamefont{Yang}},
  \bibinfo{author}{\bibfnamefont{B.}~\bibnamefont{Chen}},
  \bibinfo{author}{\bibfnamefont{K.}~\bibnamefont{Yoshimura}},
  \bibnamefont{and} \bibinfo{author}{\bibfnamefont{F.~M.}
  \bibnamefont{Grosche}}, \bibinfo{journal}{Phys. Rev. Lett.}
  \textbf{\bibinfo{volume}{109}}, \bibinfo{pages}{237008}
  (\bibinfo{year}{2012}),
  \urlprefix\url{http://link.aps.org/doi/10.1103/PhysRevLett.109.237008}.

\bibitem[{\citenamefont{Zhou et~al.}(2012)\citenamefont{Zhou, Zhang, Hong, Pan,
  Qiu, Dong, Li, and Li}}]{Zhou-2012}
\bibinfo{author}{\bibfnamefont{S.~Y.} \bibnamefont{Zhou}},
  \bibinfo{author}{\bibfnamefont{H.}~\bibnamefont{Zhang}},
  \bibinfo{author}{\bibfnamefont{X.~C.} \bibnamefont{Hong}},
  \bibinfo{author}{\bibfnamefont{B.~Y.} \bibnamefont{Pan}},
  \bibinfo{author}{\bibfnamefont{X.}~\bibnamefont{Qiu}},
  \bibinfo{author}{\bibfnamefont{W.~N.} \bibnamefont{Dong}},
  \bibinfo{author}{\bibfnamefont{X.~L.} \bibnamefont{Li}}, \bibnamefont{and}
  \bibinfo{author}{\bibfnamefont{S.~Y.} \bibnamefont{Li}},
  \bibinfo{journal}{Phys. Rev. B} \textbf{\bibinfo{volume}{86}},
  \bibinfo{pages}{064504} (\bibinfo{year}{2012}),
  \urlprefix\url{http://link.aps.org/doi/10.1103/PhysRevB.86.064504}.

\bibitem[{\citenamefont{Harada et~al.}(2007)\citenamefont{Harada, Akutagawa,
  Miyamichi, Mukuda, Kitaoka, and Akimitsu}}]{Harada-JPSJ-2007}
\bibinfo{author}{\bibfnamefont{A.}~\bibnamefont{Harada}},
  \bibinfo{author}{\bibfnamefont{S.}~\bibnamefont{Akutagawa}},
  \bibinfo{author}{\bibfnamefont{Y.}~\bibnamefont{Miyamichi}},
  \bibinfo{author}{\bibfnamefont{H.}~\bibnamefont{Mukuda}},
  \bibinfo{author}{\bibfnamefont{Y.}~\bibnamefont{Kitaoka}}, \bibnamefont{and}
  \bibinfo{author}{\bibfnamefont{J.}~\bibnamefont{Akimitsu}},
  \bibinfo{journal}{Journal of the Physical Society of Japan}
  \textbf{\bibinfo{volume}{76}}, \bibinfo{pages}{023704}
  (\bibinfo{year}{2007}), \eprint{http://dx.doi.org/10.1143/JPSJ.76.023704},
  \urlprefix\url{http://dx.doi.org/10.1143/JPSJ.76.023704}.

\bibitem[{\citenamefont{MacLaughlin}(1976)}]{maclaughlin76}
\bibinfo{author}{\bibfnamefont{D.~E.} \bibnamefont{MacLaughlin}},
  \emph{\bibinfo{title}{in Solid State Physics}} (\bibinfo{publisher}{Academic
  Press}, \bibinfo{year}{1976}).

\bibitem[{\citenamefont{Suhl et~al.}(1959)\citenamefont{Suhl, Matthias, and
  Walker}}]{suhl59}
\bibinfo{author}{\bibfnamefont{H.}~\bibnamefont{Suhl}},
  \bibinfo{author}{\bibfnamefont{B.~T.} \bibnamefont{Matthias}},
  \bibnamefont{and} \bibinfo{author}{\bibfnamefont{L.~R.}
  \bibnamefont{Walker}}, \bibinfo{journal}{Phys. Rev. Lett.}
  \textbf{\bibinfo{volume}{3}}, \bibinfo{pages}{552} (\bibinfo{year}{1959}),
  \urlprefix\url{http://link.aps.org/doi/10.1103/PhysRevLett.3.552}.

\end{thebibliography}
\end{document}